\documentclass[12pt,preprint]{revtex4-1}

\usepackage{amsmath}
\usepackage[]{graphicx}

\usepackage{times}

\newcommand{\xmax}{x_{\text{max}}}
\newcommand{\ethr}{\epsilon_{\text{thr.}}}
\newcommand{\tauhw}{\tau_{1/2}}

\begin{document}

\title{Fluctuations of the depth of maximum in extensive air showers and cross-section of p-air inelastic interaction for energy range $10^{15} - 10^{17}$~eV}

\author{S. P. Knurenko}\email{s.p.knurenko@ikfia.ysn.ru}
\author{A. Sabourov}
\affiliation{Yu. G. Shafer Institute of Cosmophysical Research and Aeronomy, 31 Lenin Ave., 677980 Yakutsk, Russia}

\begin{abstract}
  We present estimation of proton-air inelastic interaction cross-section obtained for different energy values are. Results are compared with different hadron interaction models.
\end{abstract}

\maketitle

\section{Introduction}

For more than 40 years the Yakutsk array continuously records ultra-high energy cosmic rays (UHECR) by measuring electromagnetic, muonic and Cherenkov components of extensive air showers (EAS)~\cite{Artamonov1994}. Through these measurements a lateral distribution of charged particles and muons with $\ethr \ge 1$~GeV~\cite{Dyakonov1991,Knurenko2006} was studied and longitudinal development of air showers was reconstructed by measuring the lateral distribution of Cherenkov light emitted in EAS~\cite{Dyakonov1993,Knurenko2011}.

Estimation of the depth of maximum shower development ($\xmax$) in individual showers with energy $10^{15} - 10^{19}$~eV at the Yakutsk array was performed using readings from integral and differential (tracking) Cherenkov light detectors. Integral detectors provide observation of Cherenkov photons lateral distribution in wide distance range, while differential detectors record time-base of the Cherenkov light pulse at different core distances. It follows from model calculations that characteristics such as $P = \log_{10}{Q(200)/Q(550)}$ (a relation between the density of Cherenkov light fluxes at $200$ and $550$~m from shower axis) and $\tauhw$ (the width of the Cherenkov pulse at the half-height) are sensitive to the depth of maximum shower development. Besides, the $P$ parameter depends weakly on conditions of Cherenkov light distribution in the atmosphere ($\sim 8$\,\%), but depends strongly on the accuracy of shower geometry reconstruction ($\sim 13$\,\%). The $\tauhw$ is measured by equipment that provides high precision of the Cherenkov pulse shape reproduction and thus assures better accuracy in $\xmax$ measurement ($\sim 25$~g/cm$^{-2}$). The methodology of Cherenkov light measurement and analysis is presented in a greater detail in papers~\cite{Dyakonov1991, Ivanov2007}.

\section{Methodology of the analysis}

To extract information about the parameters of interaction between the primary particle and air nuclei it is necessary to measure those characteristics of EAS which are morphologically connected to the start of nuclear cascade in air. One can test the sensitivity of EAS characteristics to cross-section of inelastic interaction with Monte-Carlo method to simulate artificial showers using some model of hadronic interactions. Such calculations were performed in multiple works. Further we will consider the methodology and algorithm for estimation of inelastic interaction cross-section involving the data on the depth of maximum EAS development ($\xmax$).

\begin{figure}
  \centering
  \includegraphics[width=0.75\textwidth, clip]{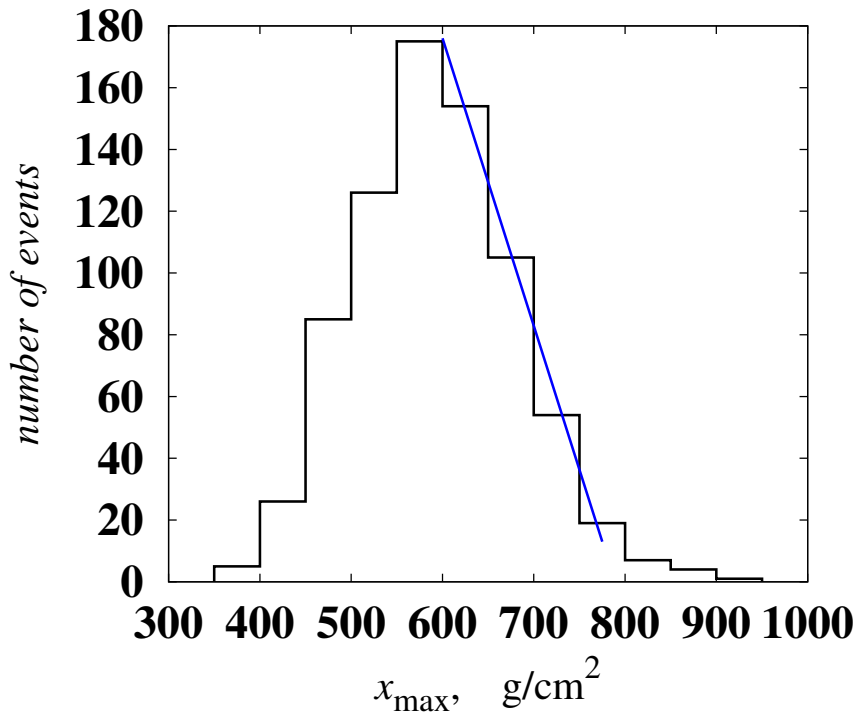}
  \caption{Fluctuations of $\xmax$ at fixed energy and the technique of $\sigma_{\text{p-air}}$ measurement by Cherenkov light data adopted at the Yakutsk array. $983$~events, $E_{0} = 2.1 \times 10^{16}$~eV, $\Lambda_{\text{exp}} = 102.6 \pm 5.2$~g/cm$^{-2}$, $\xmax = 572 \pm 6.6$~g/cm$^{-2}$, $\sigma(\xmax) = 56 \pm 3$~g/cm$^{-2}$.}
  \label{fig1a}
\end{figure}

Total cross-section of secondary particles production in EAS is one of fundamental values that characterize hadronic interactions. For showers initiated by primary protons of some energy it is possible to obtain the distribution of the depth of initial interaction ($x_{1}$):
\begin{equation}
  \frac{\mathrm{d} p}{\mathrm{d} x_{1}} = \frac{1}{\lambda_{\text{p-air}}} \cdot e^{-\frac{x_{1}}{\lambda_{\text{p-air}}}}\text{,}
  \label{eq1}
\end{equation}
where $\lambda_{\text{p-air}}$ is a mean free path of a proton before interaction in air. Mean depth of first interaction with respect to its fluctuations is directly connected to the cross-section of inelastic interaction between proton and air nucleus:
\begin{equation}
  \sigma_{\text{p-air}} = \frac{\left<m_{\text{air}}\right>}{\lambda_{\text{p-air}}}
  \label{eq2}
\end{equation}
Here, $\left<m_{\text{air}}\right>$ is mean air ``mass'', which is $\sim 14.5 m_{\text{p}} \sim 24253.01$~mb/g\,cm$^2$~\cite{Ulrich2009}. The relations (\ref{eq1}) and (\ref{eq2}) were used to estimate the inelastic interaction cross-section by experimental data obtained at the Yakutsk array. During the measurement of the $\lambda_{\text{p-air}}$ value, a mass composition of primary cosmic rays (PCR) was taken into account (contribution to generation of $\xmax$ distribution at fixed energy of other nuclei) by utilizing the points shifted by one $\sigma(\xmax)$ from the maximum of histogram (see fig.\ref{fig1a}).

\section{Experimental data}

\begin{figure}
  \centering
  \includegraphics[width=0.75\textwidth,clip]{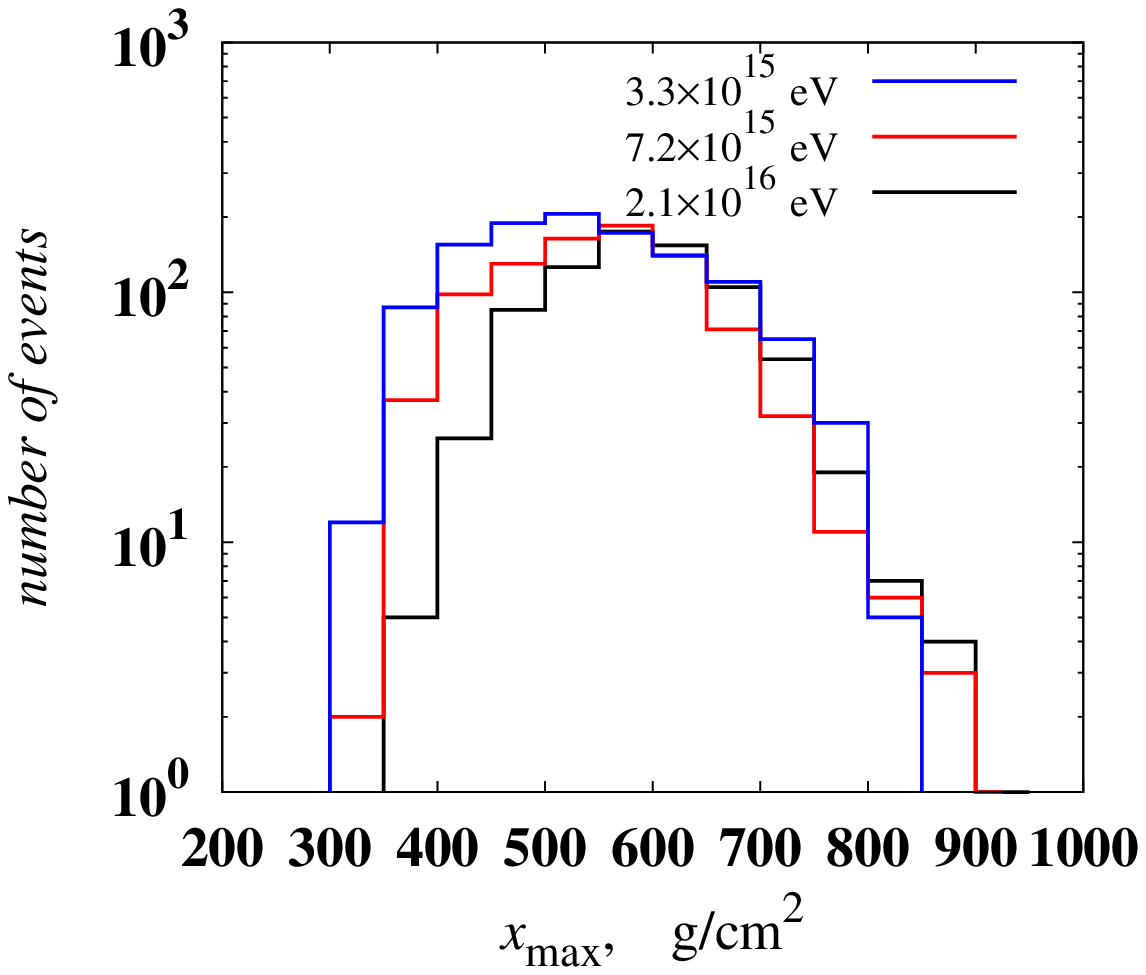}
  \caption{Initial $\xmax$ distributions, from which the $\Lambda_{\text{exp.}}$ was derived.}
  \label{fig1b}
\end{figure}

\begin{figure}[t]
  \centering
  \includegraphics[width=0.95\textwidth, clip]{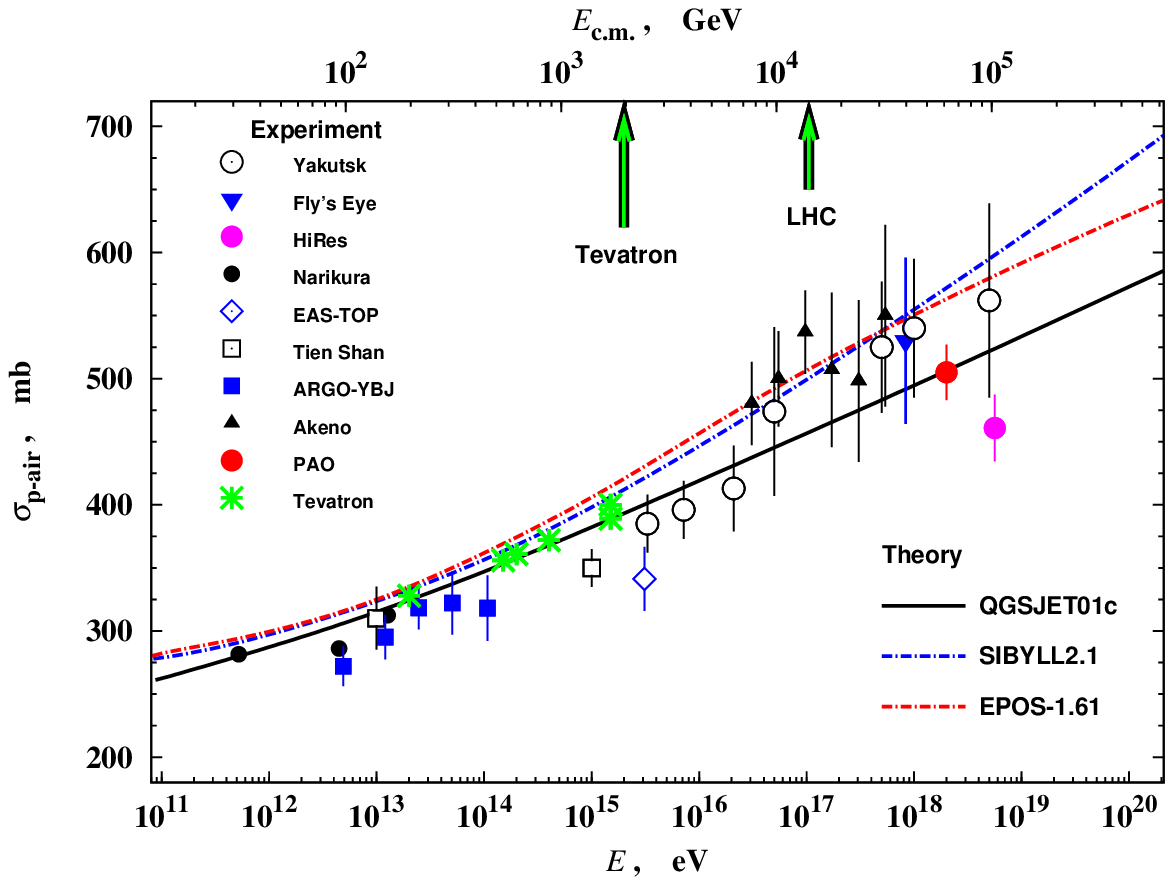}
  \caption{A comparison of current results with the other experiments and with some hadron interaction models. The original figure is taken from the work~\cite{Ulrich2009}.}
  \label{fig2}
\end{figure}

In present work we employ measured $\xmax$ distribution at fixed energy values $3.3 \times 10^{15}$~eV, $7.2 \times 10^{15}$~eV and $2.1 \times 10^{16}$~eV which were obtained from Cherenkov light data recorded by the small Cherenkov setup during more than ten years of continuous observation (see fig.\ref{fig1b}). Final results of the analysis are presented in table~\ref{tab1} and fig.\ref{fig2}.

\begin{table}
  \centering
  \caption{Cross-sections of p-air inelastic interaction}
  \label{tab1}
  \begin{tabular}{llllll}
    \hline
    $E_{0}$, eV & $\Lambda_{\text{exp.}}$ & $k_{1}$ & $k_{2}$ & $\lambda_{\text{phys.}}$, g/cm$^{-2}$ & $\sigma^{\text{prod.}}_{\text{phys.}}$, mb\\
    \hline
    $3.3 \times 10^{15}$ & $110.3 \pm 4.1$ & $1.4 \pm 0.3$ & $1.25$ & $63.0 \pm 5.2$ & $385 \pm 23$ \\
    $7.2 \times 10^{15}$ & $107.4 \pm 4.1$ & $1.4 \pm 0.3$ & $1.25$ & $61.4 \pm 5.2$ & $396 \pm 23$ \\
    $2.1 \times 10^{16}$ & $102.6 \pm 5.2$ & $1.4 \pm 0.3$ & $1.25$ & $58.7 \pm 5.2$ & $413 \pm 34$ \\
    \hline
  \end{tabular}
\end{table}

Fig.\ref{fig2} shows a compilation of the $\sigma_{\text{p-air}}(E_{0})$ data obtained in different experiments including the Yakutsk array in the energy range $10^{17} - 10^{19}$~eV (see~Tien Shan data for 1~PeV~\cite{TienShan2009}). In this work, using similar algorithm and performing the simulation of $\xmax$ measuring process with Monte Carlo method (with respect to hardware uncertainties), we obtained the $\sigma^{\text{prod.}}_{\text{exp.}}$ for energies $10^{16} - 10^{17}$~eV. New results are shown on fig.\ref{fig2}. On the same figure results of calculations with various hadron interaction models are shown~\cite{Ulrich2009}. Extrapolation of accelerator data leads to a large uncertainty at energy above $10^{18}$~eV. To a greater extent this is tied to some discrepancy between models in description of more precise accelerator data and different approaches in description of generation of secondary particles in EAS. As it is seen on fig.~\ref{fig2}, cosmic ray data together with accelerator data allow to introduce some limitations on applicability of particular hadron interaction model for description of ultra-high energy air shower development.

\section{Conclusion}

The Yakutsk EAS array is a medium sized array compared to compact and large arrays. It effectively controls the energy range from $10^{15}$~eV to $5 \times 10^{19}$~eV. In several decades of continuous observations a large amount of experimental data was accumulated. Both irregularities in cosmic ray energy spectrum are manifested in these data and the mass composition of primary particles initiating ultra-high energy air showers has been estimated using one technique. Such simultaneous analysis of the spectrum and mass composition allows one to test various hypotheses on cosmic rays origin and their propagation in the Universe. To increase the precision of this analysis it would take a choice of a single hadron interaction model from the group of models presented today. Studying the energy dependence of inelastic interaction cross-section just helps make such a choice. For example, the choice of the model for EAS development would allow experimenters to improve the precision in energy estimation and obtain the model-independent estimation of cosmic ray mass composition in ultra-high energy domain, more reliable than those adopted nowadays.

The work is supported in part by SB RAS (integral project ``Modernization of the Yakutsk array''), RFBR (grant \#11--02--00158) and the Russian Ministry of Education and Science (contract \#02.740.11.0248).

\end{document}